\documentclass[final]{svjour3}
\usepackage{graphicx}
\usepackage{rotating}
\usepackage{amssymb}
\usepackage{mathptmx}
\usepackage[numbers]{natbib}
\usepackage{amsmath}
\makeatletter
\journalname{Journal of Low Temperature Physics}

\bibpunct{[}{]}{,}{n}{}{,}

\begin{document}

\newcommand{\hdblarrow}{H\makebox[0.9ex][l]{$\downdownarrows$}-}
\title{Parallel-Plate Capacitor Titanium Nitride Kinetic Inductance Detectors for Infrared Astronomy}

\author{J. Perido\textsuperscript{1} \and P. K. Day\textsuperscript{2} \and A. D. Beyer\textsuperscript{2} \and N. F. Cothard\textsuperscript{3} \and S. Hailey-Dunsheath\textsuperscript{4} \and H. G. Leduc\textsuperscript{2} \and B. H. Eom\textsuperscript{2} \and J. Glenn\textsuperscript{3}}

\institute{Department of Astrophysics and Planetary Sciences, University of Colorado at Boulder,\\ Boulder, CO 80309, USA\\
\email{joanna.perido@colorado.edu}\\
\textsuperscript{1} University of Colorado-Boulder\\
\textsuperscript{2} NASA Jet Propulsion Laboratory (JPL)\\
\textsuperscript{3} NASA Goddard Space Flight Center\\
\textsuperscript{4} California Institute of Technology}

\maketitle

\begin{abstract}
The Balloon Experiment for Galactic INfrared Science (BEGINS) is a concept for a sub-orbital observatory that will operate from $\lambda$ = 25-250 $\mu$m to characterize dust in the vicinity of high-mass stars. The mission’s sensitivity requirements will be met by utilizing arrays of 1,840 lens-coupled, lumped-element kinetic inductance detectors (KIDs) operating at 300 mK. Each KID will consist of a titanium nitride (TiN) parallel strip absorbing inductive section and parallel plate capacitor deposited on a Silicon (Si) substrate. The parallel plate capacitor geometry allows for reduction of the pixel spacing. At the BEGINS focal plane the detectors require optical NEPs from $2\times10^{-16}$ W/$\sqrt{\textrm{Hz}}$ to $6\times10^{-17}$ W/$\sqrt{\textrm{Hz}}$ from 25-250 $\mu$m for optical loads ranging from 4 pW to 10 pW.  We present the design, optical performance and quasiparticle lifetime measurements of a prototype BEGINS KID array at 25 $\mu$m when coupled to Fresnel zone plate lenses. For our optical set up and the absorption efficiency of the KIDs, the electrical NEP requirement at 25 $\mu$m is $7.6\times10^{-17}$ W/$\sqrt{\textrm{Hz}}$ for an aborped optical power of 0.36 pW. We find that over an average of five resonators the the detectors are photon noise limited down to about 200 fW, with a limiting NEP of about $7.4\times10^{-17}$ W/$\sqrt{\textrm{Hz}}$.


\end{abstract}

\section{Introduction}
\label{sec:Intro}
The far-infrared (IR) region is rich with information needed to characterize interstellar dust. BEGINS is a concept for a sub-orbital observatory with the goal to map spectral energy distributions (SEDs) of interstellar dust in the Cygnus molecular cloud complex with a resolving power ($R=\lambda/\Delta\lambda$) of $\sim$ 7 from 25-65~$\mu$m and of 3-6 from 70-250~$\mu$m. BEGINS payload consist of a gondola which will provide the structure for all subsystems, a 50-cm aluminum Cassegrain telescope, a cryogenic instrument with a 300 mK focal plane array of detectors, and the necessary readout electronics. It requires optical NEPs from $2\times10^{-16}$ W/$\sqrt{\textrm{Hz}}$ to $6\times10^{-17}$ W/$\sqrt{\textrm{Hz}}$ from 25-250 $\mu$m, for an optical load ranging from 4 pW to 10 pW. The sensitivity requirements will be met by utilizing arrays of 1,840 lens-coupled, lumped-element KIDs. KIDs are superconducting microresonators, whose inductance changes when incident radiation is absorbed. The change in inductance produces a small shift in the resonant frequency of the microresonator, allowing for the detection of photons. Each KID will consist of a TiN parallel strip absorbing inductive section and parallel plate capacitor (PPC) deposited on a Si substrate. TiN is favorable for its high surface resistance which enables a good far-IR absorber and its tunable $T_c$ to meet sensitivity requirements \cite{noroozian2012superconducting}. The PPC is used over an IDC to attempt to decrease two-level-system (TLS) noise and decrease the fractional area taken up by the capacitor which enables a compact instrument. Our goal is to develop 25 $\mu$m BEGINS KIDs, the shortest wavelength covered by BEGINS, that meet the mission's sensitivity requirement. 



\par In this paper we will focus on the design and characterization of a 25 $\mu$m BEGINS KID prototype array. In Section \ref{sec:KIDdesign}, we discuss the BEGINS KID design and optical coupling scheme. In Section \ref{sec:DarkCharac}, we present a brief discussion on the dark characterizations of the array, providing estimates of the critical temperature and kinetic inductance fraction. In Section \ref{sec:Opt_perf}, we present the optical response of the array which consist of noise measurements and the empirical NEPs of the test device. In Section \ref{sec:Conclusion}, we conclude and discuss future work to improve the sensitivity of the BEGINS KIDs.

\section{25 $\mu$m BEGINS KIDs Design}
\label{sec:KIDdesign}
\begin{figure}[h]%
\centering
\includegraphics[width=0.7\textwidth]{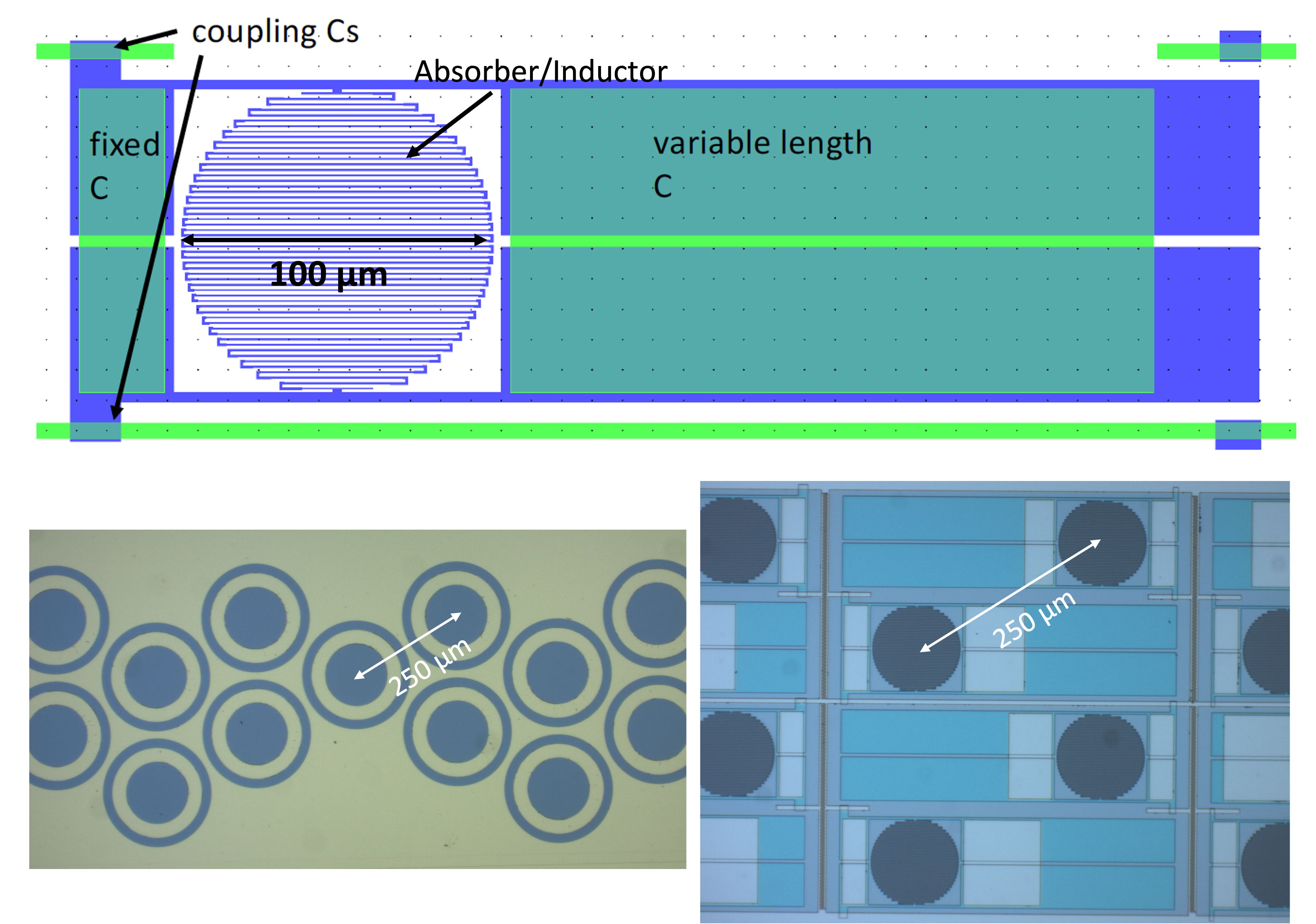}
\caption{\textit{Top}) Schematic of a BEGINS TiN KID pixel for $\lambda$= 25 µm. Purple: 50 nm TiN. The inductor and capacitor base electrodes are comprised of TiN. Green: 200-300 nm Niobium (Nb). The capacitor top electrode and coupling feedlines are compromised of Nb. 150 nm thick amorphous Silicon (aSi) is placed between the capacitor electrodes. \textit{Bottom Left}) FZP lens array layout. Blue regions are transparent zones. Green region makes up the opaque gold zones. \textit{Bottom Right}) BEGINS TiN KID array pixel layout.}
\label{fig:KIDDesign}
\end{figure}

BEGINS KIDs are lumped-element KIDs (LEKID) which consist of a resonator with a discrete inductor and capacitor. A schematic of the BEGINS LEKID is shown in Fig. \ref{fig:KIDDesign} \textit{Top}. The inductive portion is comprised of a meander in a circular envelope, with a diameter of 100 $\mu$m. The base layer (purple portion) is made of 50 nm thick sub-stoichiometric titanium nitride (TiN) with $T_c$ adjusted in the range of 1 - 1.5 K. This layer stays fixed across the array and makes up the inductors and bottom electrodes of the PPC. The top electrode of the PPC and readout lines are made of patterned 200-300 nm thick niobium (Nb, green). The dielectric layer between the top and bottom electrodes of the PPC is made of 150 nm thick Hydrogenated amorphous Si (a-Si:H). The resonant frequencies across the prototype array are set by the length of the top electrode. The designed BEGINS prototype array has 192 KIDs with a pixel pitch of 250 $\mu$m.





\par The KIDs are back-illuminated through the Si by Fresnel zone plate (FZP) lenses. FZP lenses are single layer planar structures that can easily be patterned onto the backside of the Si substrate that the KIDs are patterned on. They consist of concentric annular rings that alternate between transparent and opaque zones which focus light onto a sample through diffraction and interference \cite{wiltse1985fresnel}. The opaque zones of the FZP lenses are made of 100 nm gold (green portion of Fig. \ref{fig:KIDDesign} \textit{Bottom Left}). The focal length of each FZP lens is set to the Si substrate thickness of 625 $\mu$m and is expected to have an efficiency of 0.23. 17 out of the 192 KIDs are not coupled to an FZP lens, instead they are blocked with gold to prevent illumination. Flight devices will be equipped with full-depth microlens arrays \cite{cothard2023}. The FZP lenses however provided a quick path to optical testing of the KIDs.

\section{Prototype Array Characterization}
\label{sec:DarkCharac}
\begin{figure}[h]%
\centering
\includegraphics[width=0.8\textwidth]{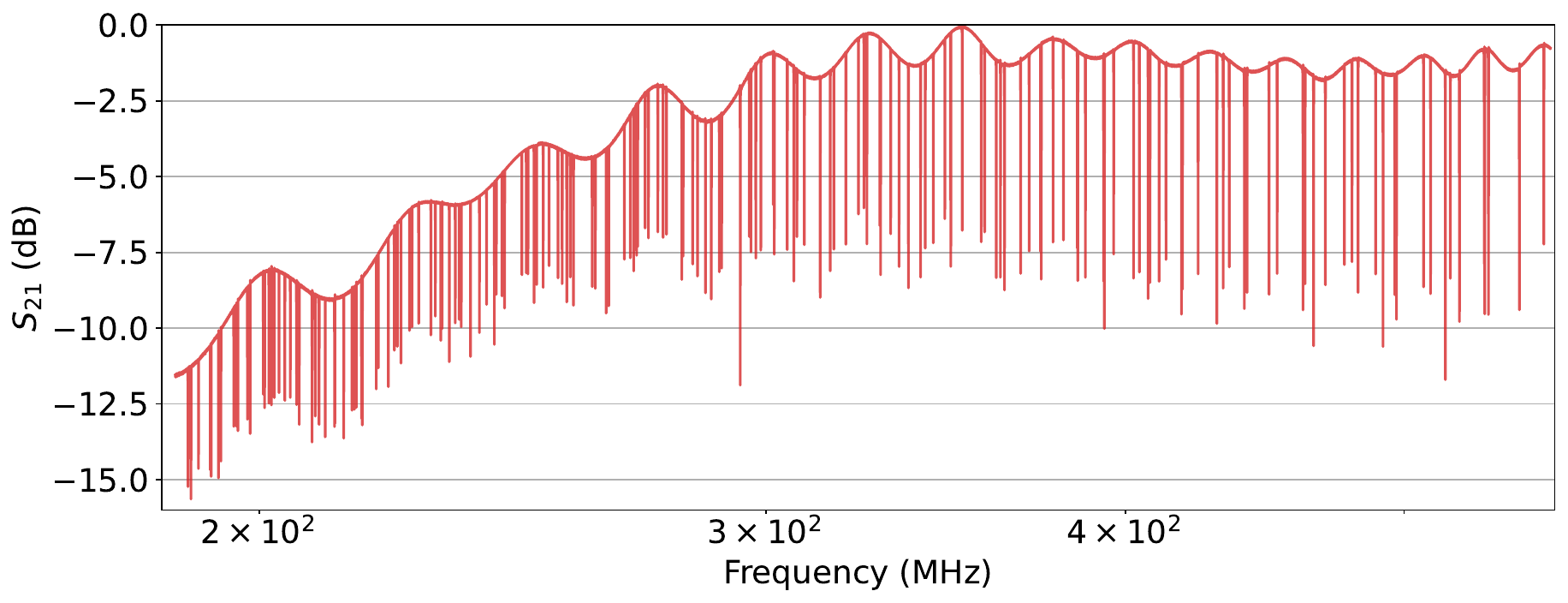}
\caption{$S_{21}$ VNA Sweep of prototype BEGINS TiN KID array. Roll-off at 300 MHz is due to room-temp amplifier.}
\label{fig:S21}
\end{figure}

The measured frequency span of the array ranged from 189-560 MHz (Fig.\ref{fig:S21}), which is close to the design span of 154-509 MHz. 181 out of the 192 resonators were identified, providing a yield of 94.3\%. The average coupling and internal quality factors of the array were $Q_c=7\times 10^{4}$ and $Q_i=7\times 10^{4}$.


\par A bath temperature sweep from 300 mK to 600 mK was performed on four resonators to estimate the critical temperature ($T_c$) and kinetic inductance fraction ($\alpha$) of the array by applying the model and fitting method described by Perido et al. \cite{perido2019}. The fit yielded an average $T_c=(1.43\pm.008)$ K and $\alpha=0.23\pm.008$ for the four resonators. The $T_c$ was within the expected design $T_c$ of 1-1.5 K. The low estimate for $\alpha$ is unusual since TiN is known to have a high $\alpha$ in the literature \cite{hubmayr2015photon}. These results could be a product of the fit. However, since we were not able to make a four-wire resistance measurement we could not verify $T_c$. We plan to make this measurement on future test devices.  
 
\section{Optical Performance Results}
\label{sec:Opt_perf}
Optical test were performed in a cryogenic testbed with a cryogenic blackbody at temperatures of 6, 40, 60, 80, 90, and 100 K. The optical set up included two band-defining filter stacks for 25 $\mu$m radiation. The measurements were made at a bath temperature of 326 mK, because the optical load of the blackbody at 100 K caused the 300 mK stage to heat up to $\sim$ 324 mK. This will result in an increase in quasiparticle generation-recombination (G-R) noise, which can increase the empirical NEP at lower optical loading where the detector is not photon-noise limited. Two different types of measurements were made to determine the empirical NEPs of the prototype array. We recorded $S_{21}$ sweeps with a VNA at each blackbody temperature to measure the fractional frequency response of the detectors relative to the lowest blackbody power (6 K). From this measurement we can estimate the responsivity ($R_x$) of the detectors. The second set of measurements were single tone noise measurements of 12 different resonators converted to power spectral densities (PSDs) in units of fractional frequency shift, $S_{xx}$. The empirical NEP of the device is calculated by, $NEP=\sqrt{S_{xx}}/R_x$.



\par
\begin{figure}[h!]
    \centering
    \includegraphics[width=1\textwidth]{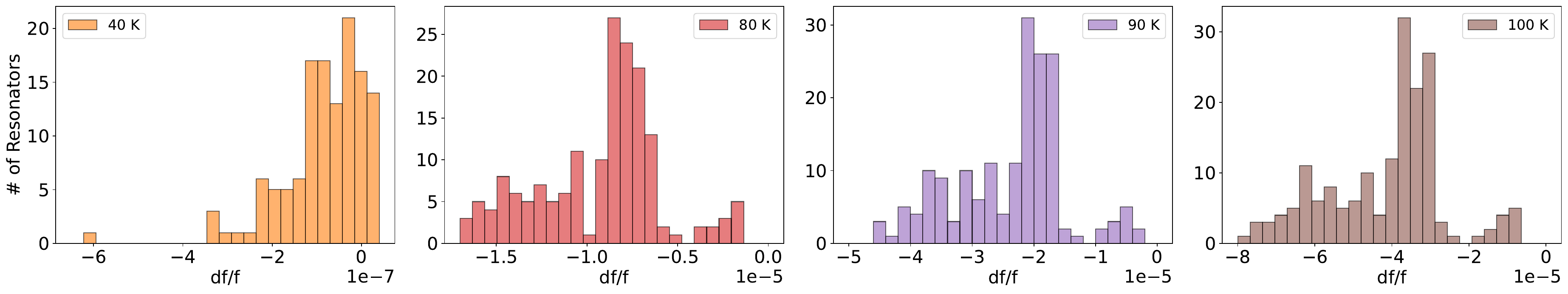}
    \caption{Histograms of BEGINS TiN KID array responses (df/f) at blackbody temperatures 40 K, 80 K, 90 K, and 100 K relative to the response at 6 K. Since there are 17/192 KIDs not coupled to a Fresnel zone plate lens we expect there to be a group of KIDs with low response due to chip heating, stray light or radiation trapped in the Si substrate. This group is seen from 80-100 K. The low response group in 80-100 K have the same 12 resonators.}
    \label{fig:BEGINS_DffHisto}
\end{figure}

 Fig. \ref{fig:BEGINS_DffHisto}, shows histograms of the resonators fractional frequency response. From 80-100 K there is a distinct group of 12 resonators with low response. These resonators may include some of those that are not coupled to FZP lenses. However, we could not confirm this because we were not able to spatially map the measured frequencies to the design frequencies. The group of resonators to the left of the low response resonators show a skewed distribution. Focusing on the response at 80 K there is a group which we referred to as the ``mid response group", with an average response of $-0.78\times10^{-5}$ from $-1.0\times10^{-5}$ to $-0.50\times10^{-5}$. The rest of the resonators in the group to the left of the mid response group are referred to as the ``high response group". The skewed distribution at 90 K and 100 K were also split between mid and high response, where resonators pertaining to each group are the same from 80-100K. We inferred that the skewed distribution may be due to radiation not coupled to the absorber by the FZP lenses scattering within the Si substrate and being absorbed by neighboring KIDs.


The hexagonal pattern of the lens array permits a detector to have 2, 3, 4 or 5 neighboring FZP lenses. Through statistical analysis we inferred that the mid response group likely consisted of KIDs with 2, 3 or 4 neighboring FZP lenses and the high response group consisted of KIDs with 5 neighboring FZP lenses. We found the mid response group estimated the response of the detectors best, so we focused on this group for the remainder of this paper. 

\subsection{Responsivities $R_{x}$}
\label{sec:R_x}
The responsivity of a KID ($R_x$) is defined as the change in fractional frequency shift due to a change in the absorbed optical power \cite{zmuidzinas2012superconducting},

\begin{equation}
    \label{eq:Rx}
    R_x = \frac{d x}{d P_{abs}}=\frac{\alpha \gamma S_2(\omega)\eta_{pb}\tau_{qp}}{4N_0\Delta_0^2V},
\end{equation}

\noindent where $d x$ is the change in fractional shift of the resonant frequency, $d P_{abs}$ is the change in absorbed optical power, $\gamma=1$ is the thin film parameter, $\eta_{pb}$ is the quasiparticle pair-breaking efficiency, $\tau_{qp}$ is the quasiparticle lifetime, $N_0$ is the single-spin density of electron states, $\Delta_0$ is the band gap energy, $V$ is the volume of the inductor, and $S_2(\omega)$ is a perturbation in the imaginary part of the conductivity of the resonator \cite{zmuidzinas2012superconducting}. From this equation a model to fit to the measurements of fractional frequency shift as a function of absorbed optical power is derived. The derivation requires the following expressions for $\tau_{qp}$ and $n_{qp}$ (the quasiparticle density),

\begin{equation}
    \label{eq:tau_qp_theory}
    \tau_{qp} = \frac{\tau_{max}}{1+\frac{n_{qp}}{n^*}}, \  n_{qp} = \sqrt{(n_{th}+n^*)^2+2(\Gamma_{opt}+\Gamma_a)\tau_{max}n^*/V}-n^*,
\end{equation}

   

\noindent where $n^*$ is the crossover density at which the observed quasiparticle lifetime saturates to $\tau_{max}$ \cite{zmuidzinas2012superconducting}, $\Gamma_{opt}$ is the generation rate
due to absorption of optical photons, and $\Gamma_{a}$ is the generation rate due to the absorption of readout photons. Under the assumption that $\Gamma_a \ll \Gamma_{opt}$ Eq. \ref{eq:Rx} becomes,

\begin{equation}
    \label{eq:deltax_dnqp_2}
    R_{x} = \frac{d x}{d P_{abs}} = \frac{\alpha \gamma S_2(\omega) \eta_{pb}}{4 N_0 \Delta_0^2 V} \left(1+\frac{n_{th}}{n^*}\right)^{-1} \left[1+\frac{2 \eta_{pb} P_{abs} \tau_{max}}{\Delta_0 n^*V \left(1+\frac{n_{th}}{n^*}\right)^2}\right]^{-1/2}
    =\frac{R_{x,0}}{\sqrt{1+\frac{P_{abs}}{P_0}}} , 
\end{equation}



\noindent where $R_{x,0} = \frac{\alpha \gamma S_2(\omega) \eta_{pb} \tau_{max}}{4 N_0 \Delta_0^2 V} \left(1+\frac{n_{th}}{n^*}\right)^{-1} \textrm{ and } P_0 = \frac{\Delta_0 n^*V}{2 \eta_{pb} \tau_{max}}\left(1+\frac{n_{th}}{n^*}\right)^2$. The parameter $R_{x,0}$ is the responsitvity in the limit that the absorbed optical power is zero and $P_0$. By integrating the simplified form of Eq. \ref{eq:deltax_dnqp_2} we obtain a model for the fractional frequency shift ($x$) as a function of absorbed optical power ($P_{abs}$),  



\begin{equation}
\label{eq:x_of_P_model_2}
    x = 2R_{x,0}P_0{\sqrt{1+\frac{P_{abs}}{P_0}}} + C,
\end{equation}

\noindent where C is a constant that arises from integration. This yields a model with a three parameter fit for $R_{x,0}$, $P_0$, and C. The fit parameters $R_{x,0}$ and $P_0$ plugged into Eq. \ref{eq:deltax_dnqp_2}, provide $R_x$ as a function on $P_{abs}$. 

Next, we calculate the absorbed blackbody optical power, $P_{BB,abs}$, by taking the following integral over a given wavelength range,

\begin{equation}
\label{eq:BEGINS_Pinc}
    P_{BB,abs} = A_{pixel}\Omega\eta_{opt}\eta_{abs}\int_{\lambda_i}^{\lambda_f}\frac{B(\lambda, T)}{2}F_\lambda\, d\lambda,
\end{equation}

\noindent where $B(\lambda, T)$ is Planck's law for blackbody radiation, $A_{pixel}$ is the area of an FZP lens, $\Omega$ is the solid angle subtended by the blackbody onto the FZP lens, $\eta_{abs}=0.77$ is the efficiency at which the detector absorbs optical power, $F_\lambda$ is the transmission of the filter stacks as a function of wavelength, and $\eta_{opt}$ is the optical efficiency that accounts for reflections off the Si-vacuum interface at the FZP lenses (=0.7) and the FZP efficiency of 0.23. The division of two arises becasue the absorber is only sensitive to one polarization.





\begin{figure}[h!]
    \centering
    \includegraphics[width=.85\textwidth]{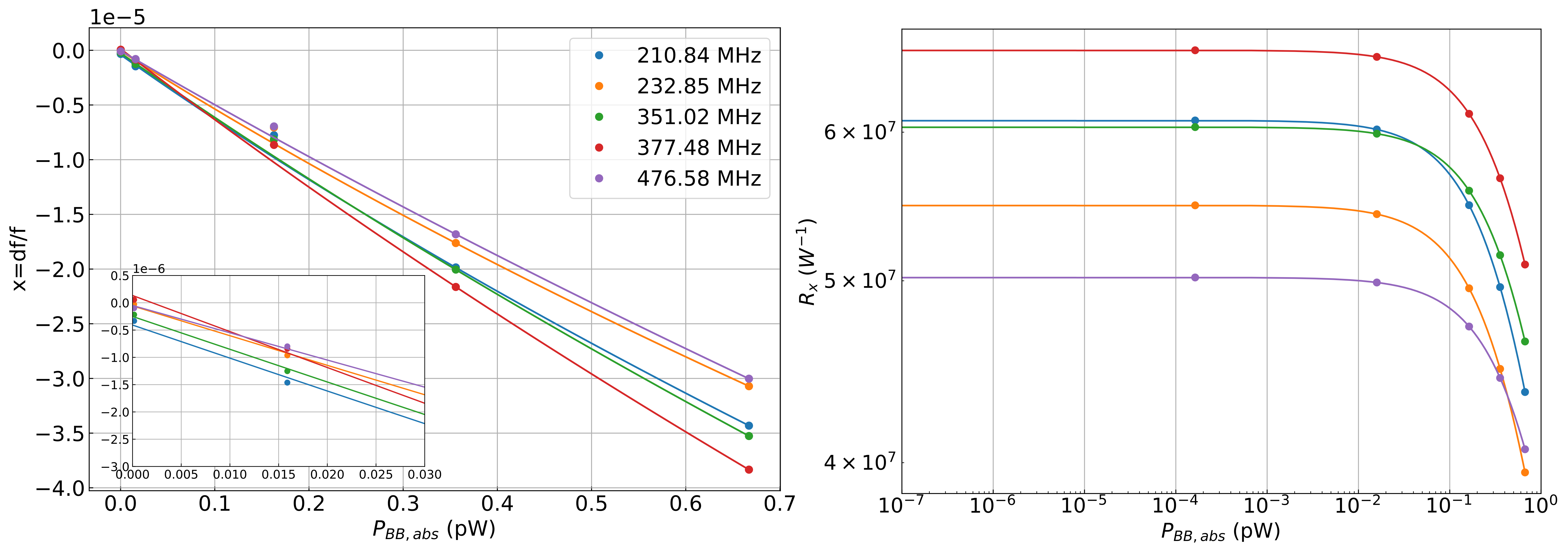}
    \caption{\textit{Left}) Fractional frequency response as a function of absorbed blackbody power for five resonators. Dots: Data. Lines: Model from Eq. \ref{eq:x_of_P_model_2} fit to data. \textit{Right}) Responsivity of the five resonators as a function of absorbed blackbody power.}
    \label{fig:BEGINS_Rx}
\end{figure}

Fig. \ref{fig:BEGINS_Rx} \textit{Left}, shows the measurements (dots) and the fits to the model derived above (lines) for five resonators in the mid response group. The response at 0.16 pW is lower than expected when compared to the trend followed by the rest of the data, so this data point was excluded from the fits. This may have been due to an unstable blackbody that had not settled to 80 K. The model agrees well with the measurements, indicating that it describes the response to absorbed optical power well. The estimated parameters, $R_{x,0}$ and $P_0$, from the fits were used to calculate $R_x$ as a function of absorbed power (Fig. \ref{fig:BEGINS_Rx} \textit{Right}). The scatter in response may be due to the amount of absorbed stray light which depends on the number of neighboring FZP lenses or location of the detectors along the chip. Detectors at the center of the chip directly below the blackbody aperture will have a larger optical efficiency. 


\subsection{Noise Measurements \& NEPs}
\begin{figure}[h!]
    \centering
    \includegraphics[width=.95\textwidth]{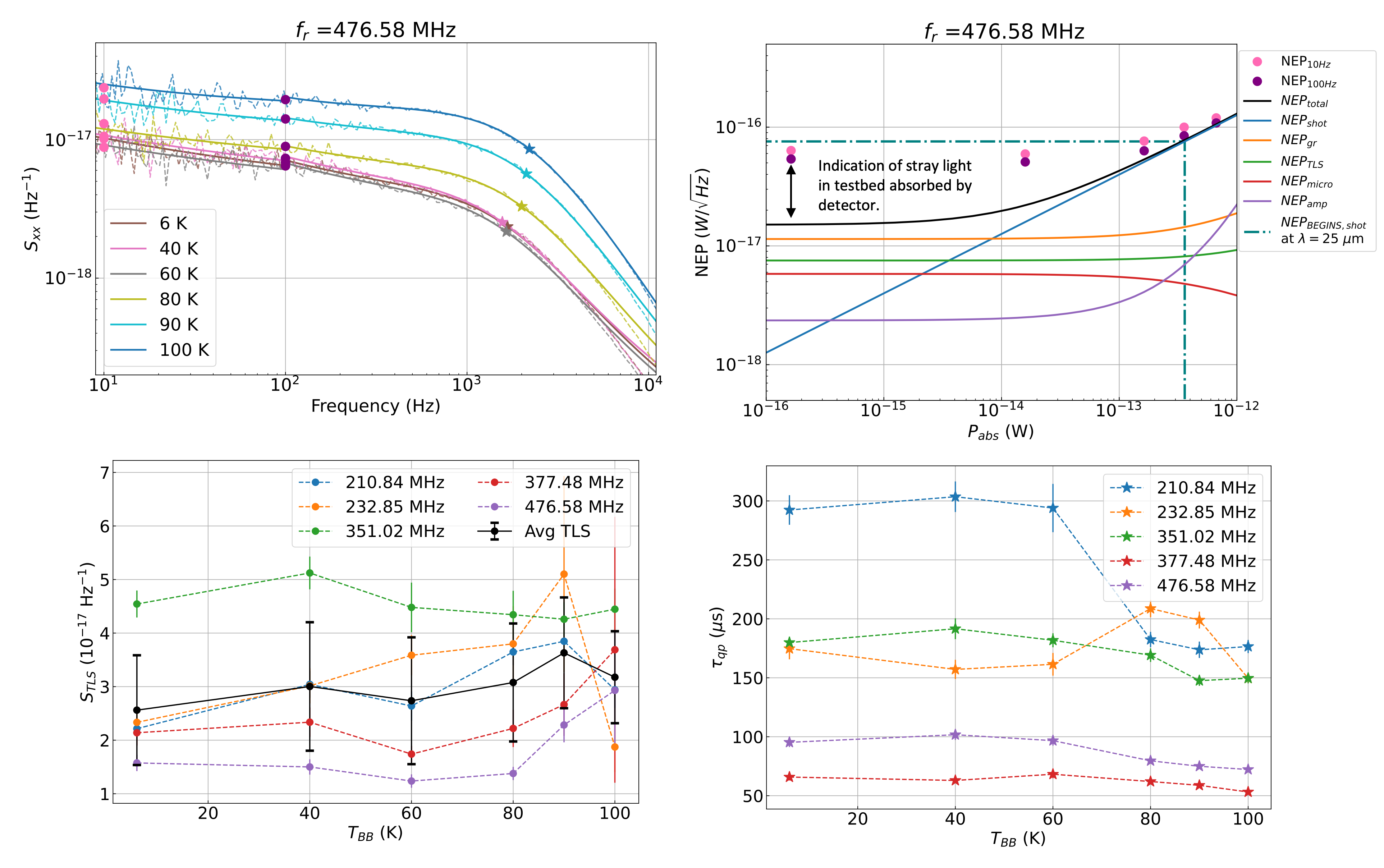} 
    \caption{\textit{Top Left}) Phase noise PSDs ($S_{xx}$) of a TiN KID with increasing cryogenic blackbody temp (increasing absorbed optical load). Dashed lines: Measured phase noise. Solid lines: Fits to $S_{xx}$. Stars: Estimated quasiparticle lifetimes ($\tau_{qp}$). \textit{Top Right}) Dots: Empirical NEPs at $S_{xx}$ = 10 Hz and 100 Hz for resonator with $f_r$ = 476.58 MHz. Teal dash-dotted line: BEGINS NEP requirement at 25 $\mu$m. Black line: Total theoretical NEP. Other solid lines represent the NEP contributions due to microwave power, generation-recombination noise, amplifier noise, and TLS noise. \textit{Bottom}) Estimates of the $S_{xx}$ noise fit parameters $S_{TLS}$ and $\tau_{qp}$ for five resonators. The solid black line is the average TLS noise level over all five resonators.}
    \label{fig:BEGINS_NEP}
\end{figure}

Fig. \ref{fig:BEGINS_NEP} \textit{Top Left} shows the measured frequency noise PSDs (dashed lines) at each blackbody temperature for one detector, $f_r = 476.58$ MHz. The PSDs show 1/f noise up to the roll-off above 1 kHz, which is dominated by TLS noise. As the optical load increases the PSDs begin to flatten and increase in amplitude, due to an increase in white noise which originates from photon noise and G-R noise. The photon noise is expected to dominate at high loading. The roll-off seen above 1 kHz is due to the $\tau_{qp}$ of the detectors. 

\par To estimate the TLS noise and $\tau_{qp}$ we fit the PSDs to the following model, 
\begin{equation}
    \label{eq:PSD_model}
    S_{xx}=\frac{S_{WN}}{1+(2\pi \nu \tau_{qp})^2}+S_{TLS}f^{-n}.
\end{equation}

\noindent The first term is a Lorentzian with a roll-off at $\tau_{qp}$ and amplitude ($S_{WN}$) that depends on the white noise level. The second term is used to fit for the 1/f noise which has an amplitude that depends on the TLS noise ($S_{TLS}$), and n is the spectral index of the TLS noise \cite{noroozian2009two,kumar2008submillimeter,gao2008semiempirical}. ${S_{WN}}$, $S_{TLS}$, $\tau_{qp}$, and $n$ are all free parameters when fitting the model. The model has a hard time capturing $\tau_{qp}$ at low optical loads when the TLS noise dominates, so we fit the PSDs in sections. To estimate $S_{TLS}$ the PSDs are fit from 10 Hz to 100 Hz. To capture the roll-off properly the PSDs are fit from 100 Hz to 10 kHz. The fits are shown as solid lines in Fig. \ref{fig:BEGINS_NEP} \textit{Top Left}. The stars represent the fit for the roll-off frequency. Fig. \ref{fig:BEGINS_NEP} \textit{Bottom} shows the estimated $S_{TLS}$ and $\tau_{qp}$ as a function of blackbody temperature for all five mid response resonators. At 10 Hz $S_{TLS}$ varies from $1\times10^{-17}$ to $5 \times10{-17}$ Hz\textsuperscript{-1} which is comparable to values seen in aluminum and niobium KIDs with IDCs by Z. Pan et al. \cite{pan2023noise}. They studied how TLS noise varies with IDC gap width. At a drive power comparable to the drive power of the BEGINS resonators ($>$-93 dBm) the TLS noise in their resonators varied from $1.5\times10^{-17}$ to $4\times10^{-17}$ 1/Hz. This shows that PPCs are a promising design choice for KIDs that allow for smaller pixels. $\tau_{qp}$, varies from $\sim$50-300 $\mu$s which is within a common range seen for different TiN KID devices in the literature \cite{diener2012design,leduc2010titanium, wheeler2019millimeter}.

\par The empirical NEPs were calculated at 10 Hz (pink dots) and 100 Hz (dark purple dots) for the five resonators as a function of absorbed blackbody power and were photon-noise limited down to about 200 fW. Fig. \ref{fig:BEGINS_NEP} \textit{Top Right} shows the results for the resonator with  $fr = 476.58$ MHz. The solid blue line represents the cryogenic blackbody photon noise NEP, where $NEP_{shot}=\sqrt{2P_{BB,abs}h\nu_{ph}(1+n_0)}$. The teal dotted dashed line represents the expected BEGINS photon noise NEP at 25 $\mu$m, which is expected to be $7.58\times10^{-17}$ W/$\sqrt{\textrm{Hz}}$ for an absorbed power of 0.36 pW. The BEGINS absorbed power was calculated by taking the expected BEGINS incident optical load (4 pW) at 25 $\mu$m and applying the necessary efficiencies ($\eta_{opt}$ and $\eta_{abs}$) and the division by two from absorption of one polarization. The average electrical NEP of the five resonators at the expected BEGINS absorbed power were NEP\textsubscript{10 Hz} = $9.2\times10^{-17}$ W/$\sqrt{\textrm{Hz}}$ and NEP\textsubscript{100 Hz} = $7.5\times10^{-17}$ W/$\sqrt{\textrm{Hz}}$. At 10 Hz the NEP is a factor of 1.2 above the BEGINS requirement, which can be decreased by increasing $R_x$ of the TiN KIDs. This can be done by increasing $\tau_{qp}$ or decreasing $\Delta_0$ or the volume of the inductor. As we have shown $\tau_{qp}$ of TiN varies, so the best approach is changing the stoichiometry of the TiN such that $\Delta_0$ is smaller or reducing the volume of the inductor.  


The empirical NEPs were compared to the total theoretical NEP, shown as a black line in Fig. \ref{fig:BEGINS_NEP} \textit{Top Right}. The total NEP of the test device was calculated by adding all NEP contributions in quadrature, $\ NEP_{tot}^2 = NEP_{TLS}^2 + NEP_{amp}^2 + NEP_{photon}^2 + NEP_{GR}^2 + NEP_{micro}^2$, where $NEP_{amp}$ is due to amplifier noise, and $NEP_{micro}$ is due to readout power quasiparticle generation noise \cite{zmuidzinas2012superconducting, gao2008physics}. Below $\sim$ 0.25 pW $NEP_{tot}$ is not photon-noise limited and is dominated by G-R and TLS noise. The measured NEPs show a similar behavior but saturate to $5.1\times10^{-17}$ W/$\sqrt{\textrm{Hz}}$ at powers below 0.1 pW, which is larger than the total theoretical NEP. This could be due to stray light in the cryostat, stray light from neighboring FZP lenses not captured by the model or the model is underestimating $NEP_{tot}$.



\section{Conclusion}
\label{sec:Conclusion}
We have successfully designed and tested a prototype 25 $\mu$m BEGINS TiN KID array that was back-illuminated by an FZP lens array. The goal of this work was to determine if the BEGINS NEP requirement was achievable in a lab set up within the expected modulation rate of the BEGINS instrument $\sim$1-10 Hz. At 10 Hz we measured an empirical NEP of $9.2\times10^{-17}$ W/$\sqrt{\textrm{Hz}}$, a factor of 1.2 above the requirement of $7.6\times10^{-17}$ W/$\sqrt{\textrm{Hz}}$ for an absorbed optical power of 0.36 pW. We also found that the average $S_{TLS}=3\times10^{-17}$ Hz \textsuperscript{-1} and that $\tau_{qp}$ varied from 50-300 $\mu$s for the device. To improve the sensitivity we have designed and fabricated a new 25 $\mu$m BEGINS TiN KID array with a smaller inductor volume and higher absorption efficiency. This array will be bonded to a Si microlens array. Optical test have been performed on a KID array successfully bonded to a microlens array, where the response was not affected by stray light from neighboring lenses \cite{cothard2023}. This will improve the NEPs and decrease the scatter in its value across the array. We believe these changes will improve the empirical NEPs and achieve the BEGINS NEP requirements at 10 Hz. 


\textbf{Acknowledgments} Portions of this research were carried out at the Jet Propulsion Laboratory, California
214 Institute of Technology, under a contract with NASA (80NM0018D0004). This work was also supported by internal grants at NASA GSFC. Joanna Perido was supported by the NASA Future Investigators in NASA Earth and Space Science. Nicholas F. Cothard was supported by the NASA Postdoctoral Program Fellowship at NASA GSFC.



\end{document}